\documentclass[conference]{IEEEtran}
\IEEEoverridecommandlockouts

\usepackage{cite}
\usepackage{amsmath,amssymb,amsfonts}
\usepackage{algorithmic}
\usepackage{graphicx}
\usepackage{textcomp}
\usepackage{xcolor}
\usepackage{subcaption}
\usepackage{url}

\def\BibTeX{{\rm B\kern-.05em{\sc i\kern-.025em b}\kern-.08em
    T\kern-.1667em\lower.7ex\hbox{E}\kern-.125emX}}
\begin{document}

\title{A Research and Development Portfolio of GNN Centric Malware Detection, Explainability, and Dataset Curation\\

}

\author{\IEEEauthorblockN{1\textsuperscript{st} Hossein Shokouhinejad}
\IEEEauthorblockA{\textit{Faculty of Computer Science} \\
\textit{University of New Brunswick}\\
Fredericton, Canada \\
hossein.shokouhinejad@unb.ca}
\and
\IEEEauthorblockN{2\textsuperscript{nd} Griffin Higgins}
\IEEEauthorblockA{\textit{Faculty of Computer Science} \\
\textit{University of New Brunswick}\\
Fredericton, Canada \\
Griffin.Higgins@unb.ca}
\and
\IEEEauthorblockN{3\textsuperscript{rd} Roozbeh Razavi-Far}
\IEEEauthorblockA{\textit{Faculty of Computer Science} \\
\textit{University of New Brunswick}\\
Fredericton, Canada \\
roozbeh.razavi-far@unb.ca}
\and
\IEEEauthorblockN{4\textsuperscript{th} Ali A Ghorbani}
\IEEEauthorblockA{\textit{Faculty of Computer Science} \\
\textit{University of New Brunswick}\\
Fredericton, Canada \\
ghorbani@unb.ca}
}

\maketitle

\begin{abstract}
Graph Neural Networks (GNNs) have become an effective tool for malware detection by capturing program execution through graph-structured representations. However, important challenges remain regarding scalability, interpretability, and the availability of reliable datasets. This paper brings together six related studies that collectively address these issues. The portfolio begins with a survey of graph-based malware detection and explainability, then advances to new graph reduction methods, integrated reduction–learning approaches, and investigations into the consistency of explanations. It also introduces dual explanation techniques based on subgraph matching and develops ensemble-based models with attention-guided stacked GNNs to improve interpretability. In parallel, curated datasets of control flow graphs are released to support reproducibility and enable future research. Together, these contributions form a coherent line of research that strengthens GNN-based malware detection by enhancing efficiency, increasing transparency, and providing solid experimental foundations.
\end{abstract}


\section{Introduction}
Malware continues to pose a critical challenge to cybersecurity, with modern threats demonstrating increasing sophistication and diversity. Traditional signature-based detection methods struggle to keep pace with these evolving techniques, underscoring the need for more adaptable and robust solutions. Graph-based representations of program behavior, such as control flow graphs (CFGs) and system call graphs, have emerged as an effective means of capturing structural and semantic information that is often overlooked by conventional approaches. In this context, Graph Neural Networks (GNNs) provide a powerful framework for modeling such data, enabling the identification of complex patterns that distinguish malicious from benign software \cite{MalGNE, 14_ZHEN2025110524,DawnGNN,GAT3}.

Despite their potential, the application of GNNs to malware detection faces several unresolved challenges. The size and complexity of program graphs raise concerns about scalability, as training and inference can become prohibitively resource-intensive. At the same time, the black-box nature of GNN models limits interpretability, which is especially problematic in security domains where analysts require clear and reliable explanations to validate decisions. Furthermore, progress in this area is slowed by the scarcity of curated, large-scale datasets that capture realistic malware behavior and support reproducible experimentation. These limitations highlight the need for systematic efforts that address efficiency, explainability, and data availability in a unified manner.
\begin{figure*}
    \centering
    \includegraphics[width=\linewidth]{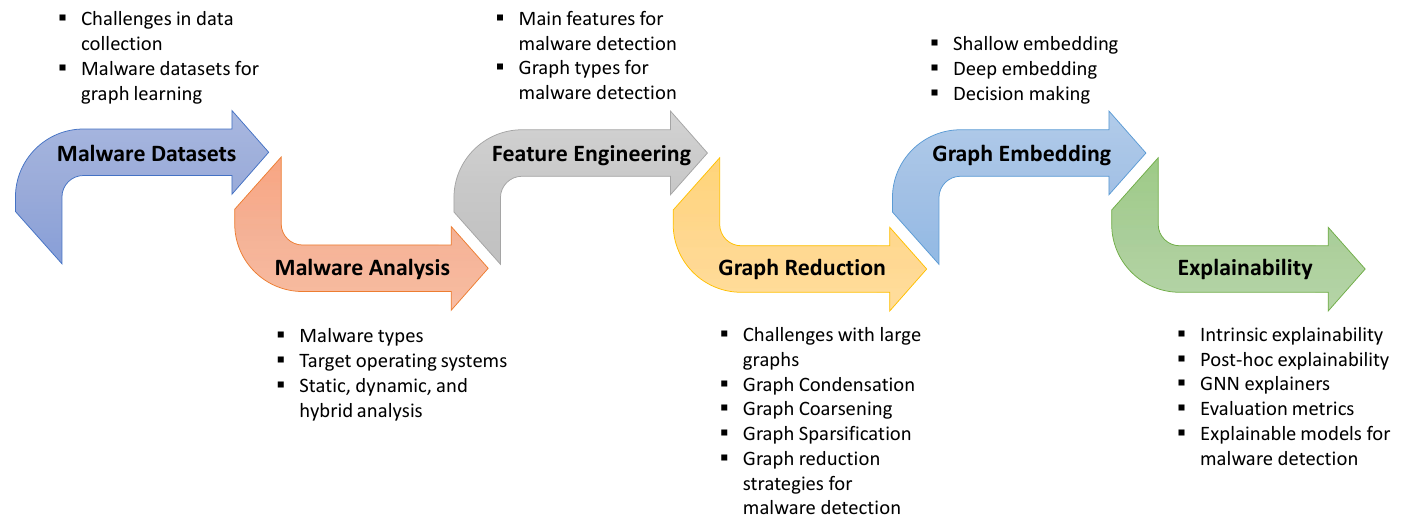}
    \caption{Roadmap of graph-based malware detection \cite{CIC3}, showing the link between datasets, analysis, feature engineering, graph reduction, embedding, and explainability.}
    \label{fig:roadmap}
\end{figure*}
This paper presents a research and development portfolio that tackles these challenges through six interconnected studies. The starting point is a survey that not only synthesizes advances in graph-based malware detection and explainability but also reviews emerging graph reduction approaches, establishing a foundation for subsequent developments. Building on this groundwork, novel reduction strategies are proposed to improve the scalability of GNNs on large program graphs, followed by methods that integrate reduction and learning for more effective malware classification. The portfolio then shifts toward interpretability, first by examining the consistency of GNN explanations and later by introducing dual explanation techniques based on subgraph matching to provide verifiable evidence for model outputs. The final step introduces an ensemble framework with attention-guided stacked GNNs, designed to enhance predictive performance while offering richer interpretability. Complementing these algorithmic advances, curated datasets of CFGs extracted from Portable Executable (PE) files are released to enable reproducibility and support future research. Collectively, these studies form a coherent progression of contributions that advance the field of GNN-centric malware detection.

Beyond the individual studies, this portfolio demonstrates how a sequence of connected research efforts can be used to address the central challenges of applying GNNs to malware detection. The collective results show that scalability, interpretability, and reproducibility can be advanced in parallel, leading to approaches that are both technically effective and suitable for practical deployment. By bringing together developments in graph reduction, explanation methods, ensemble modeling, and dataset design, the work emphasizes that progress in this field depends on treating malware detection as a dynamic and integrated research ecosystem in which algorithmic design, interpretability, and resource availability strengthen one another.

The remainder of this paper is organized as follows. Section \ref{survey} provides the survey foundation, outlining recent advances in graph-based malware detection and explainability while identifying open challenges. Section \ref{reduction} discusses graph reduction methods and their role in improving efficiency. Section \ref{explainability} examines explainability with a particular focus on explanation consistency and subgraph matching. Section \ref{ensemble} presents ensemble learning with attention-guided stacked GNNs as a step toward advanced interpretability. Section \ref{data} highlights contributions in dataset curation, particularly the release of CFGs extracted from PE files. Finally, Section \ref{conclusion} concludes the paper by summarizing the portfolio and pointing to future research directions.

\section{Survey Foundation: Mapping the Landscape}
\label{survey}
The foundation of this research portfolio is established through a comprehensive survey on graph-based malware detection and explainability \cite{CIC3}. The survey provides an extensive overview of the growing role of graph learning techniques in cybersecurity, emphasizing how GNNs can capture structural and relational information embedded in malware behavior. By framing malware analysis as a graph learning problem, the survey highlights the shift from traditional feature-based approaches to advanced methods capable of modeling complex program interactions such as control flows, function calls, and API dependencies.

A central contribution of the survey lies in its systematic review of malware datasets, analysis techniques, and feature engineering strategies \cite{CIC3}. It discusses the challenges of acquiring diverse and representative datasets, particularly for benign software, and underscores the importance of graph-aware data sources to support reproducible research. The survey further examines static, dynamic, and hybrid malware analysis methods, clarifying how each contributes unique insights into program behavior. By doing so, it positions graph learning as a unifying framework capable of integrating diverse perspectives into a single analytical pipeline.

The survey also explores graph reduction and embedding approaches in depth, addressing the scalability challenge posed by large program graphs \cite{CIC3}. Reduction methods such as sparsification, condensation, and coarsening are presented as strategies for simplifying complex graphs while retaining critical structural information. These are complemented by embedding techniques, both shallow and deep, which transform graph data into lower-dimensional representations suitable for GNN training. Together, these components provide the methodological basis for scalable and efficient malware detection, ensuring that advanced models remain computationally practical without sacrificing performance.
\begin{figure*}[h]
    \centering
    \begin{subfigure}[b]{0.32\textwidth}
        \centering
        \includegraphics[width=\linewidth]{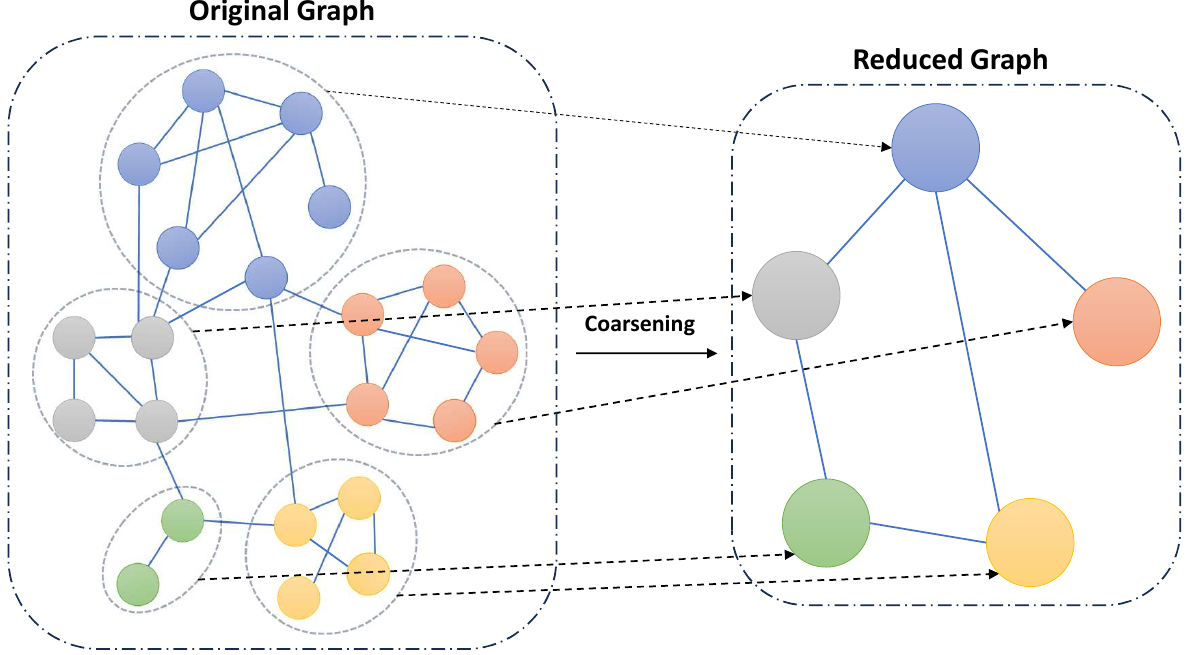}
        \caption{Coarsening}
        \label{fig:sub1}
    \end{subfigure}
    \hfill
    \begin{subfigure}[b]{0.32\textwidth}
        \centering
        \includegraphics[width=\linewidth]{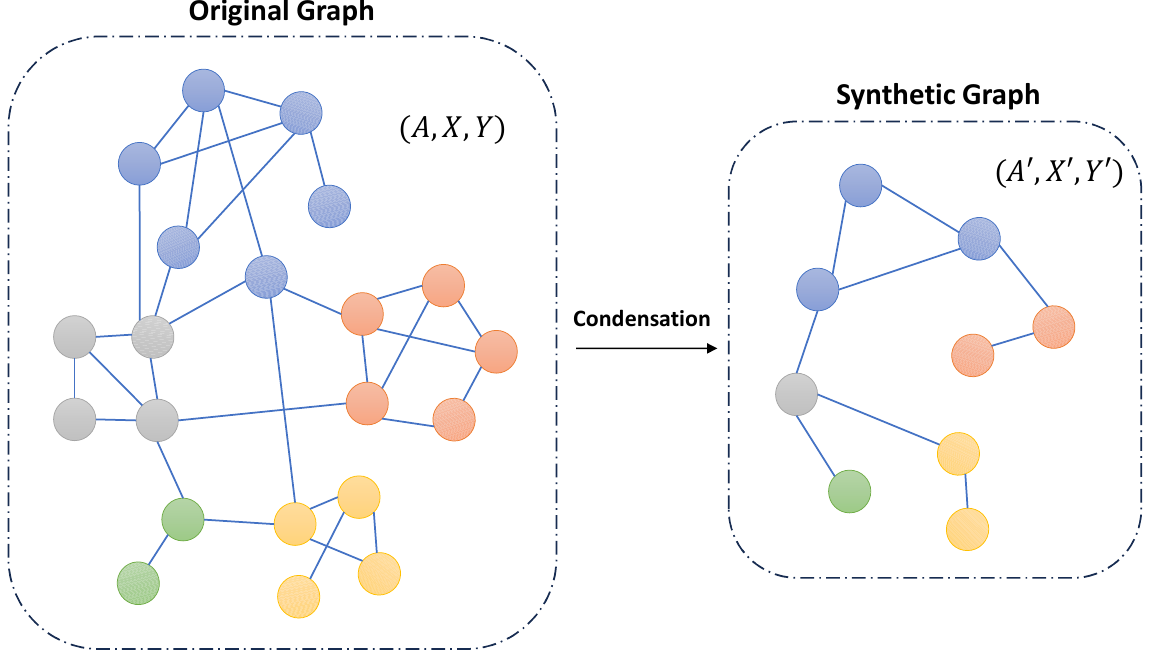}
        \caption{Condensation}
        \label{fig:sub2}
    \end{subfigure}
    \hfill
    \begin{subfigure}[b]{0.32\textwidth}
        \centering
        \includegraphics[width=\linewidth]{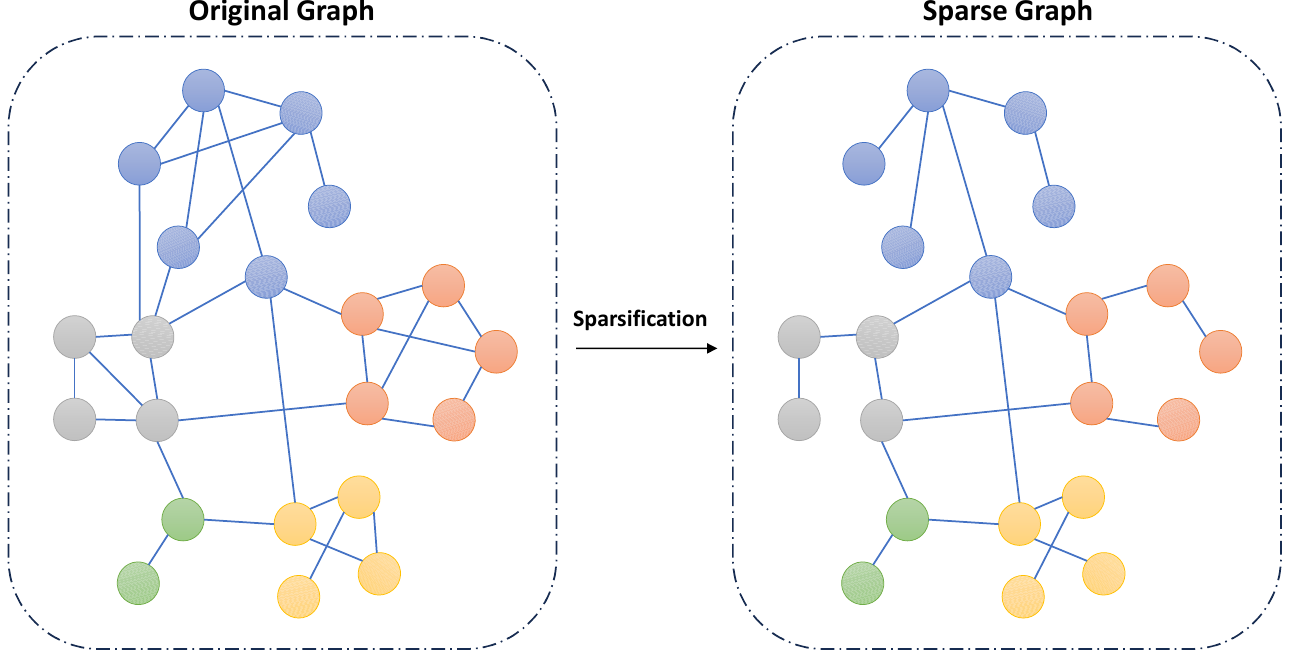}
        \caption{Sparsification}
        \label{fig:sub3}
    \end{subfigure}
    
    \caption{Overview of graph reduction techniques, including coarsening, condensation, and sparsification \cite{CIC3}.}
    \label{fig:reduction_methods}
\end{figure*}

Finally, the survey dedicates substantial attention to explainability as an indispensable element of GNN-based malware detection \cite{CIC3}. It categorizes explainability approaches into intrinsic and post-hoc methods, highlighting how GNN explainers can identify critical nodes, edges, and subgraphs that drive classification outcomes. This focus on interpretability is essential in cybersecurity, where model transparency directly impacts analyst trust and regulatory compliance. By integrating discussions on datasets, analysis techniques, reduction methods, embeddings, and explainability, the survey establishes a coherent roadmap for future research and sets the stage for the more specialized contributions developed in the remainder of this portfolio. The overall flow of these interconnected components is illustrated in Figure~\ref{fig:roadmap}, which presents a roadmap adapted from the survey \cite{CIC3}.

\section{Graph Reduction and Efficiency}
\label{reduction}

The size and complexity of program graphs extracted from PE files, particularly CFGs, present significant challenges for GNN-based malware detection. These graphs often contain thousands of nodes and edges, including redundant structures, peripheral blocks, and noisy components that contribute little to classification accuracy. Processing such graphs directly increases computational cost, slows down training and inference, and reduces the interpretability of model decisions. Graph reduction has therefore become an essential pre-processing strategy to simplify these structures while retaining their most informative features. Three main families of methods have been studied in this regard: graph coarsening, graph condensation, and graph sparsification. Coarsening merges nodes and edges to create smaller, approximate versions of the original graph; condensation synthesizes highly compact graphs that preserve statistical properties of the original; and sparsification removes less significant nodes or edges to produce a reduced but still structurally faithful graph. Figure~\ref{fig:reduction_methods} illustrates these three broad categories, providing the conceptual foundation for the reduction approaches explored in this portfolio \cite{CIC3}.

One of the main contributions to this domain is Node-Centric Pruning (NCP), a novel sparsification technique specifically designed for malware detection graphs \cite{Node_centric}. Unlike conventional edge-based methods, NCP adopts a node-first perspective, categorizing nodes into three groups: Nexus Nodes, Connector Nodes, and Sparse Nodes. Nexus Nodes are identified through an exhaustive analysis of walks of fixed length $L$, which quantify each node’s connectivity and reachability within the graph. Nodes with insufficient connectivity are further classified as either Connector Nodes, linked to Nexus Nodes but less central, or Sparse Nodes, which are peripheral and weakly connected. Sparse Nodes are pruned in the first stage to reduce graph noise. In the second stage, Connector Nodes undergo refinement based on Jaccard similarity with Nexus Nodes. Those with low similarity are removed to ensure only structurally meaningful nodes remain. Through this two-step strategy, NCP achieves significant reductions in graph size while maintaining the topological properties critical for downstream GNN performance. Experiments showed that NCP consistently outperforms state-of-the-art sparsification techniques such as Walk Index Sparsification (WIS), delivering improved scalability and stable classification accuracy across malware graph datasets.

Complementary to NCP, another study proposed an integrated framework that combines graph reduction with learning and explainability modules \cite{CIC1}. This framework employs multiple pruning strategies, including Leaf Prune, Component Prune, k-core decomposition, and WIS, to reduce CFGs and function call graphs (FCGs) prior to classification. By tailoring pruning to the structural properties of program graphs, the framework reduces training and inference costs without compromising predictive performance. What distinguishes this work is the explicit integration of explainability. After reduction and classification, GNNExplainer \cite{GNNExplainer} is applied to extract the most influential subgraphs driving detection outcomes. This ensures that the framework not only achieves computational efficiency but also produces explanations that are more concise and interpretable due to the reduced complexity of the underlying graphs. Evaluations demonstrated that pruning substantially improves efficiency and helps analysts focus on a clearer subset of program structures when interpreting results.

Taken together, these contributions establish graph reduction as a cornerstone of efficient and interpretable malware detection. NCP introduces a principled, node-centric pruning paradigm that balances reduction with structural preservation, while the integrated framework illustrates how reduction and explainability can be unified into a single pipeline. Both approaches demonstrate that scalability and trustworthiness need not be opposing goals. With well-designed reduction methods, GNN-based malware detection can achieve high accuracy, reduced computational demands, and enhanced interpretability.
\begin{figure*}
    \centering
    \includegraphics[width=\linewidth]{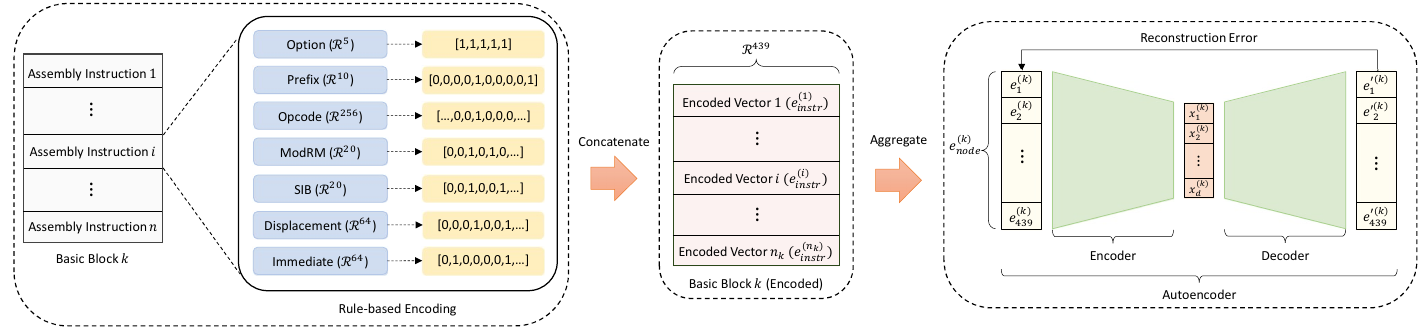}
    \caption{Node feature embedding process, where raw assembly instructions from PE file control flow graphs are converted into fixed-length vectors and compressed into low-dimensional embeddings \cite{CIC2}.}
    \label{fig:node_embedding}
\end{figure*}

\section{Explainability and Consistency}
\label{explainability}

A key dimension of this research portfolio is the pursuit of reliable and interpretable explanations for GNN-based malware detection. While GNNs provide strong predictive performance on program graphs such as CFGs, their black-box nature limits trust in high-stakes security settings. Two complementary contributions address this issue: a framework for evaluating and improving the consistency of GNN explanations \cite{CIC2}, and a dual explanation approach that aligns detected patterns with known prototypes through subgraph matching \cite{Dual}.

The first study introduces a dynamic malware detection framework that constructs CFGs from PE files and employs a hybrid node feature embedding process \cite{CIC2}. Each node (basic block) is represented by its assembly instructions, which are first encoded into a high-dimensional 439-bit vector capturing fields such as prefix, opcode, ModRM, SIB, displacement, and immediate values. To manage dimensionality and improve stability, these vectors are then compressed into 64-dimensional embeddings using an autoencoder. This process, illustrated in Figure~\ref{fig:node_embedding}, provides compact and expressive node features suitable for downstream GNN training. By combining structural graph information with semantically rich embeddings, the framework ensures that the GNN has both accurate and meaningful input representations for malware detection.

Based on this foundation, the framework evaluates multiple state-of-the-art explainers including GNNExplainer, PGExplainer \cite{PGExplainer}, and CaptumExplainer (with Integrated Gradients, Saliency, and Guided Backpropagation) \cite{Captum}. To address the variability of explanations across methods, the study introduces the RankFusion explainer, an aggregation strategy that fuses edge rankings from two top-performing explainers. This fusion enhances stability and reduces noise in the extracted explanatory subgraphs. In addition, a novel Greedy Edge-wise Composition (GEC) algorithm is proposed to construct more coherent and connected subgraphs, improving the readability and structural fidelity of explanations. The effectiveness of these techniques is evaluated through metrics such as accuracy, fidelity, and consistency, demonstrating that RankFusion and GEC together yield explanations that are both stable under perturbations and faithful to the model’s predictions.

\begin{figure*}[h]
    \centering
    \includegraphics[width=0.8\linewidth]{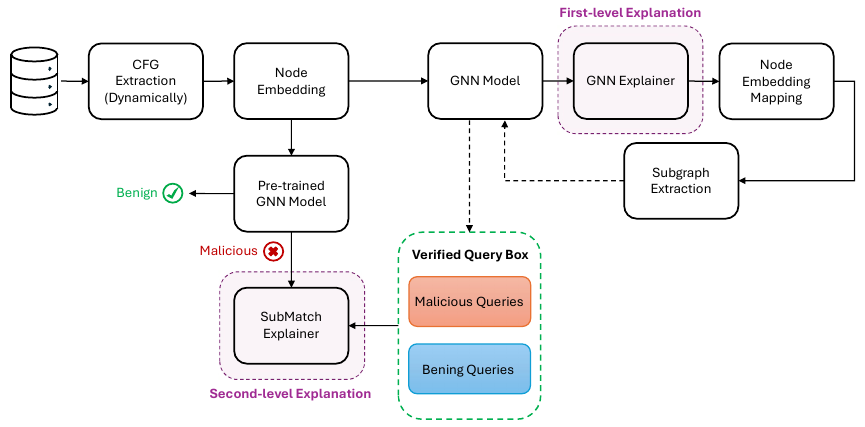}
    \caption{Dual explanation framework for malware detection, combining a base GNN explainer with a subgraph matching module (SubMatch) to connect local explanations \cite{Dual}.}
    \label{fig:dual_framework}
\end{figure*}

\begin{figure}[h]
    \centering
    \includegraphics[width=\linewidth]{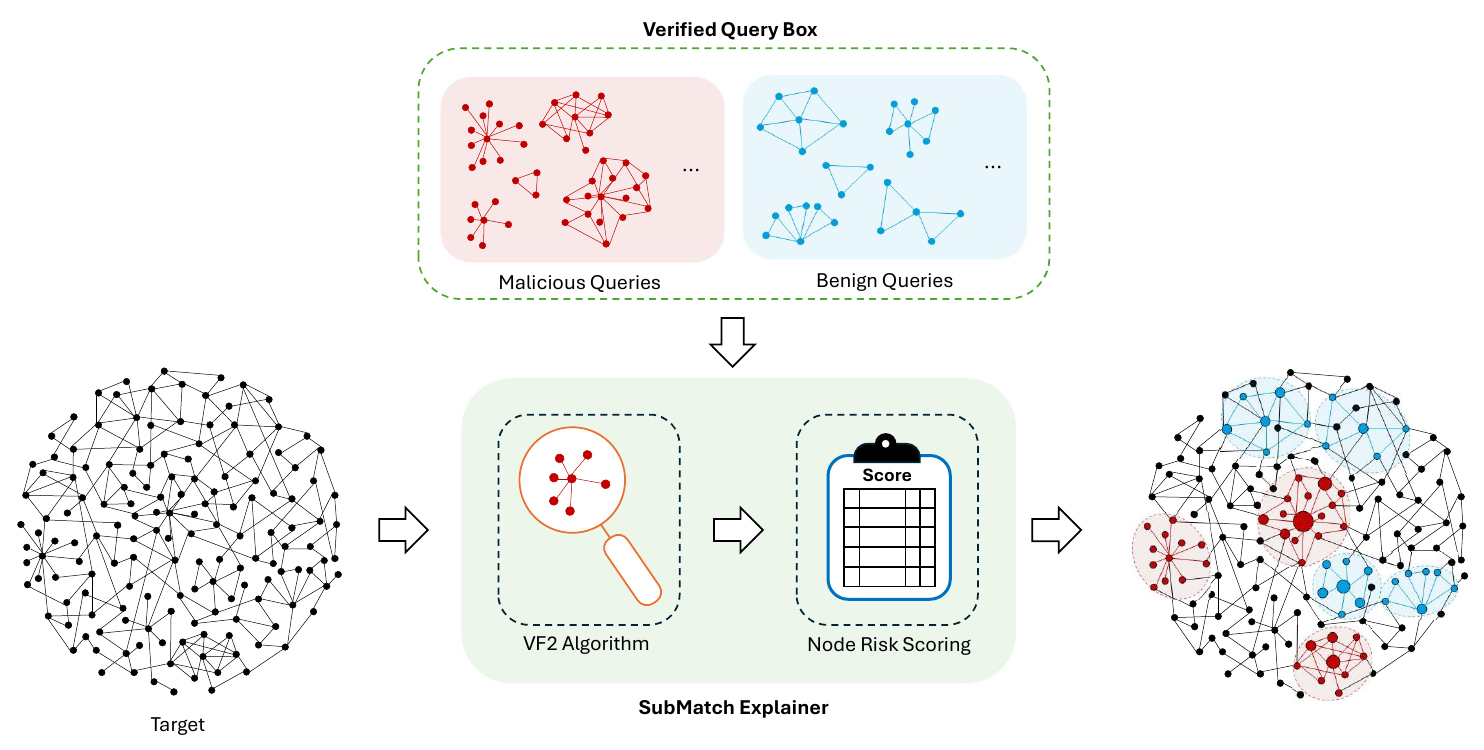}
    \caption{SubMatch explainer using subgraph matching to highlight relevant malicious (red) and benign (blue) regions within a target CFG \cite{Dual}.}
    \label{fig:submatch}
\end{figure}

The second contribution introduces a dual explainability framework that augments standard GNN explainers with a prototype-driven interpretation layer \cite{Dual}. After a GNN explainer identifies candidate subgraphs, they are verified against the classifier and stored as trusted benign or malicious prototypes in a curated query box. During testing, only samples predicted as malicious undergo this second-level explanation, where the SubMatch explainer applies subgraph matching (using algorithms such as VF2) to align regions of the target CFG with the stored prototypes. Figures~\ref{fig:dual_framework} and \ref{fig:submatch} illustrate this process. The SubMatch explainer assigns interpretable scores to nodes based on their association with verified malicious or benign subgraphs, enabling a fine-grained view of program regions that drive classification. Importantly, this approach not only highlights important subgraphs but also anchors them to behaviorally meaningful prototypes, enhancing the analyst’s ability to interpret the decision in terms of known malicious or benign behaviors.

Together, these studies push the boundary of GNN explainability in malware detection by addressing both stability and semantic grounding. The first framework ensures that explanations are consistent, reliable, and structurally coherent across different explainers and sparsity levels. The second adds a new dimension by linking explanatory subgraphs to verified behavioral prototypes, thereby moving from abstract feature attributions toward behavior-aligned explanations. By combining these advances, the portfolio strengthens the trustworthiness of GNN-based malware detection, ensuring that model outputs are not only accurate but also interpretable and actionable for cybersecurity practitioners.

\section{Ensemble Learning and Advanced Explainability}
\label{ensemble}

Single GNN models have proven effective in capturing the structural dependencies of CFGs, yet they are limited by their individual learning biases and may struggle to generalize against diverse or evasive malware. To address this challenge, an ensemble framework was developed that leverages stacking with attention-guided aggregation, combining the strengths of multiple GNNs while simultaneously enhancing interpretability \cite{Ensemble}.

The framework begins with the dynamic extraction of CFGs from PE files, ensuring that runtime behaviors such as indirect jumps and dynamically loaded code are captured. For classification, the framework employs multiple diverse GNNs as base learners, including GCN, GIN, and GAT, each characterized by a distinct message-passing mechanism. These base models capture complementary aspects of program structure, producing prediction outputs that are then passed to a meta-learner. The meta-learner is implemented as an attention-enhanced multilayer perceptron (MLP), which aggregates base model outputs and assigns attention weights to quantify the contribution of each learner. This not only improves detection accuracy but also introduces model-level interpretability by indicating which GNN architectures are most influential in the final decision.

\begin{figure}[h]
    \centering
    \includegraphics[width=\linewidth]{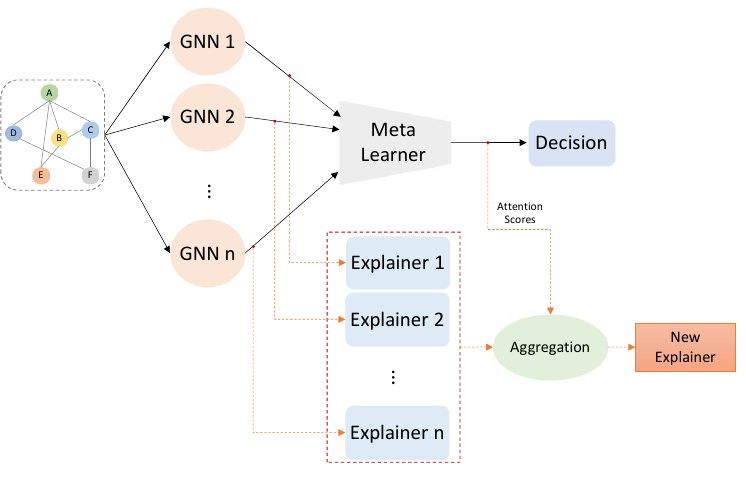}
    \caption{Attention-guided stacking ensemble for malware detection, combining diverse GNN base learners with an attention-based meta-learner and ensemble-aware explanation.}
    \label{fig:ensemble_framework}
\end{figure}

A central innovation of this study lies in its ensemble-aware explanation method. Conventional post-hoc explainers such as Integrated Gradients (IG) and Guided Backpropagation (GBP) are applied to individual base learners to identify influential edges and subgraphs. These edge-level importance scores are then aggregated according to the attention weights generated by the meta-learner, producing explanations that are aligned with the ensemble’s final prediction. This process, illustrated in Figure~\ref{fig:ensemble_framework}, enables analysts to see not only which substructures within the CFG drive classification, but also how different GNN models collectively contribute to the detection decision.

\section{Dataset Curation}
\label{data}

An important aspect of this research portfolio focuses on addressing an open challenge in graph-based Windows malware research and development through the publication of program graphs. However, acquiring binary samples before extracting such graphs is especially challenging. This is because in the benign setting proprietary consumer-facing operating systems (e.g., Windows) often enforce strong copyright protections. This is especially true for system binaries and those found in application stores where benign binaries often reside. Furthermore, the publication of malicious samples may come with legal liability for researchers that limit their publicity. This is due to the potential harm that may be inflicted on the public if samples are misused or abused \cite{CIC3}. Instead, researchers wishing to publish meaningful data and assist other malware researchers often publish features collected from binaries in order to understandably avoid such liability. However, the overall utility of such data is decreased since such highly specific features cannot be used to derive other more complex features generally. For example, the entropy of a given sample cannot be used to derive its respective CFG. Despite this, some datasets do exist with raw binaries such as DikeDataset \cite{dikedataset}, Blue Hexagon Open Dataset for Malware AnalysiS (BODMAS) \cite{yang2021bodmas}, and PE Malware Machine Learning Dataset (PMMLD) \cite{practicalsecurity2024pe}. However, using samples in these datasets often requires disarming samples before they can be used to avoid accidental execution. In some datasets a formal request must also be made to obtain the binaries. These datasets can then be filtered to match a particilar program type (e.g., Windows x86-64 PE). Afterwords, a binary analysis tool, such as angr \cite{shoshitaishvili2016state}, can be used to extract the CFG and FCG either statically or dynamically.

\subsection{CIC-SGG-2024}
As part of \cite{CIC1}, CFGs and FCGs of samples in the previously mentioned datasetes, represented as a project object in angr, are extracted. These objects contain rich information and meta-data about the analyzed binary as well as the CFG and FCG structures. Furthermore, the embedded versions of the same graphs, stored as PyTorch Geometric (PyG) graphs, used for training GNN models are also published. Specifically, the embedding techniques Function Name Embedding (FNE) and Assembly Embedding (AE) are used for FCGs and CFGs respectively to embed nodes. Additionally, an explanation for each sample, represented as an importance mask over nodes are provided as graph objects from the Networkx Python library. Importantly, two main audience groups are recognized for these samples, namely graph learning specialists and malware researchers. This also motivates their separation of attribute, embedding, and explanation graphs.

\subsection{CIC-DGG-2025}
In subsequent work \cite{CIC2}, dynamically generated CFG graphs for the same set of samples are also extracted. Such samples are much more computationally difficult to generate dynamically and require significant memory resources. However, these graphs tend to be smaller than their statically generated counterparts and generally comprise a single weakly connected component. A similar publication structure as CIC-SGG-2024 is used, though more explanation examples per the various explanation algorithms are provided that also consider edge importance weights.

\section{Conclusion}
\label{conclusion}

This portfolio demonstrates how GNNs can be advanced to meet the challenges of malware detection. It begins with a survey that mapped open problems in graph learning and explainability, followed by contributions on graph reduction for scalability, integrated pruning–explainability frameworks, and studies on the consistency of explanations. Further developments introduced dual explanation techniques based on subgraph matching and an ensemble learning approach that improved both accuracy and interpretability. Alongside these methods, curated datasets of CFG and FCG graphs were released to strengthen reproducibility in graph-based Windows malware research.

Together, these contributions outline a coherent trajectory that advances efficiency, interpretability, and reproducibility in GNN-centric malware detection. Future work may extend these directions by exploring reduction methods for adversarial robustness and broadening dataset coverage. This integration of methodological innovation with dataset publication provides a foundation for building scalable and trustworthy malware detection systems.

\bibliographystyle{ieeetr}
\bibliography{Main}

\end{document}